\documentclass[aps,showpacs,superscriptaddress,prb,twocolumn]{revtex4-1}

\usepackage[english]{babel}
\usepackage[ansinew]{inputenc} 
\usepackage[T1]{fontenc}	
\usepackage[]{graphicx} 
\usepackage{amsmath} 
\usepackage{dsfont} 
\usepackage{pgf} 
\usepackage[colorlinks=true,allcolors=black]{hyperref}

\newcommand{\bra}[1]{\ensuremath{\left\langle #1\right|}}
\newcommand{\ket}[1]{\ensuremath{\left|#1\right\rangle}}
\newcommand{\braket}[2]{\ensuremath{\left\langle #1\vphantom{#2}\right.\left|\vphantom{#1}#2\right\rangle}}
\DeclareMathOperator{\Tr}{Tr}

\begin{document}
\graphicspath{{pictures/}}
\title{Excitation- and state-transfer through spin chains in the presence of spatially correlated noise}

\author{Jan Jeske}
\affiliation{Chemical and Quantum Physics, School of Applied Sciences, RMIT University, Melbourne, 3001, Australia}
\author{Nicolas \surname{Vogt}}
\affiliation{Institut f\"ur Theorie der Kondensierten Materie,  \\ Karlsruhe Institute of Technology, D-76128 Karlsruhe, Germany}
\affiliation{DFG-Center for Functional Nanostructures (CFN), Karlsruhe Institute of Technology, D-76128 Karlsruhe, Germany}
\author{Jared H. Cole}
\affiliation{Chemical and Quantum Physics, School of Applied Sciences, RMIT University, Melbourne, 3001, Australia}

\begin{abstract}
We investigate the influence of environmental noise on spin networks and spin chains. In addition to the common model of an independent bath for each spin in the system we also consider noise with a finite spatial correlation length. We present the emergence of new dynamics and decoherence-free subspaces with increasing correlation length for both dephasing and dissipating environments.  This leads to relaxation blocking of one spin by uncoupled surrounding spins. We then consider perfect state transfer through a spin chain in the presence of decoherence and discuss the dependence of the transfer quality on spatial noise correlation length. We identify qualitatively different features for dephasing and dissipative environments in spin-transfer problems.
\end{abstract}

\maketitle

The transfer of a quantum state is an important component for quantum technology. While transfer via photons in optic fibre enables high-speed communication for long-range communication and cryptography there has also been a large interest in short-distance transfer via (pseudo-) spin chains\cite{Bose2007, Bose2003, Osborne2004, Albanese2004,Christandl2004, Christandl2005, Li2005, Shi2005, Yung2005, Burgarth2005, Burgarth2005_randcoupl, Wojcik2005, Fitzsimmons2006, Greentree2006, Kostak2007, Jafarizadeh2008, Gualdi2008}. Many promising quantum technologies, such as optical lattices\cite{Mandel2003} and arrays of quantum dots\cite{Kane1998, Loss1998}, rely on such transport. Furthermore, a quantum mechanically very similar mechanism is the transfer of excitation energy in light-harvesting complexes in the context of photosynthesis\cite{Plenio2008, Rebentrost2009, Hoyer2010, Marais2013}. 

Within the many configurations for transport in spin networks, a linear spin chain transversely coupled with a particular spatially varying coupling strength has been found to provide perfect state transfer from one end to the other\cite{Christandl2004}. We will focus on this case of perfect-state-transfer as it provides well defined analytical solutions. However, our results are more general and apply to other spin-network problems and spin-wave theory in general.

While the effects of decoherence in particular photosynthetic systems have been studied in depth\cite{Plenio2008, Rebentrost2009, Hoyer2010, Marais2013}, studies of environmental noise on perfect state transfer are limited to spatially uncorrelated noise\cite{Burgarth2006,Cai2006,Hu2009,Hu2010} or noise correlations between repeated transfers through the same chain\cite{Macchiavello2002, Arshed2006, Bayat2008, Benenti2009, D'Arrigo2012}. Here we investigate comprehensively the effects of decoherence on excitation and state transfer. Particularly we assign a characteristic spatial correlation length $\xi$ to the environmental noise and display our results as continuous functions of $\xi \in [0,\infty)$. There has been experimental evidence for such spatially correlated noise in ion traps\cite{Chwalla2007,Monz2011}. Generally, in spin chain systems with short nearest-neighbour distances one can expect non-zero spatial correlations in the environmental noise on the system length scales. 

We begin by discussing very generally the effects of spatially correlated decoherence in systems of several two level subsystems (spin-1/2, qubit, etc.), without defining the system parameters such as interqubit coupling or dimensionality specifically. 
Following this, we will consider the particular effects on perfect state transfers in spin chains.

\section{Spatially correlated effects in a system of several spins}
\label{sec spatially correlated effects}
First we investigate spatially correlated noise for a general spin system described by the Hamiltonian,
$
H_s=\sum_{j=1}^n \omega_q \sigma_z^{(j)} + H_c
$
, where $H_c$ can contain general coupling terms between the spins. These results are applicable to any spin network or qubit array and give an intuition for the analysis of any particular system. We model decoherence with Bloch-Redfield equations which derive directly from the interaction between system and environment. This interaction consists of longitudinal coupling terms (leading to dephasing) and transversal coupling terms (leading to relaxation),
$
H_{int}= \sum_{j=1}^N v_j\,\sigma_z^{(j)}\, B_\parallel^{(j)} + \sum_{j=1}^N  \nu_j\,\sigma_x^{(j)}\, B_\perp^{(j)}
$
. In the secular approximation, based on $\omega_q \gg H_c$, the two coupling types couple to separate baths and we will discuss them separately. The Bloch-Redfield equations then read for a low-temperature environment ($\hbar=1$):
\begin{align}
&\dot \rho = i [\rho,H_s] + \dots\\
&+ \sum_{j,k} v_j v_k C_\parallel(0,|x_j-x_k|) \left( \sigma_z^{(k)} \rho \sigma_z^{(j)} - \frac{1}{2} \{ \sigma_z^{(j)} \sigma_z^{(k)}, \rho\} \right) \nonumber\\
&+ \sum_{j,k} \nu_j \nu_k C_\perp(2 \omega_q,|x_j-x_k|) \left( \sigma_-^{(k)} \rho \sigma_+^{(j)} - \frac{1}{2} \{ \sigma_+^{(j)} \sigma_-^{(k)}, \rho\} \right)\nonumber
\end{align}
The environmental spectral functions 
$
C_\parallel(\omega, |x_j-x_k|)=\int_{-\infty}^{\infty} e^{i\omega \tau} \langle \tilde B_\parallel^{(j)}(\tau,x_j) \tilde B_\parallel^{(k)}(0,x_k)\rangle
$
and
$
C_\perp(\omega, |x_j-x_k|)=\int_{-\infty}^{\infty} e^{i\omega \tau} \langle \tilde B_\perp^{(j)}(\tau,x_j) \tilde B_\perp^{(k)}(0,x_k)\rangle
$ occur naturally in the formalism and set the spatial and temporal correlations of the environment. For details see ref.~\onlinecite{Jeske2013formalism}

\subsection{Spatially correlated dephasing}
\label{sec spatially correlated dephasing}
For $n$ qubits in an uncorrelated environment the dephasing rate between two states is proportional to the number $n_f$ of flipped qubits between the two states. In a perfectly correlated environment however the dephasing rate between two states with a difference of $n_e$ excitations is proportional to $n_e^2$ and $n_f$ is irrelevant\cite{Jeske2013formalism, Breuerbook}. Therefore the dephasing rate between states with equal excitation number is reduced to zero when the noise correlation length increases well beyond the qubits' separation. In other words each subspace of states with equal numbers of excitations becomes a decoherence-free subspace. On the other hand for states such as the GHZ state $\ket{111...} + \ket{000...})/\sqrt{2}$ which have $n_f = n_e$ the dephasing rate increases enormously in spatially correlated environments.

Taking four qubits as an example, an off-diagonal density matrix element of the form \ket{0011}\bra{1100} will decay with rate $\Gamma = n_f \gamma = 4\gamma$ for $\xi \rightarrow 0$ and as $\Gamma = n_e^2 \gamma = 0$ for $\xi \rightarrow \infty$, where $\gamma$ is the corresponding single qubit dephasing rate. In contrast, the coherence \ket{0000}\bra{1111} which also decays as $\Gamma = n_f \gamma = 4\gamma$ for $\xi \rightarrow 0$, will decay as $\Gamma = n_e^2 \gamma = 16 \gamma$ for $\xi \rightarrow \infty$, i.e.~the rate increases proportional to system size for long correlation lengths.

\subsection{Spatially correlated relaxation}
\label{sec spatially correlated relaxation}
When qubits couple to a bath via transversal coupling, one of the effects of correlated noise is analogous to the well-known super- and sub-radiance \cite{Dicke1954} of an atomic gas. We will give an analysis of the underlying effects with a focus on the single-excitation subspace, which is essential for excitation transfer in spin chains. We refer to ``relaxation'' solely as the loss of energy to the environment and ``excitation gain'' as the opposite.

\begin{figure}
\centering
\includegraphics[]{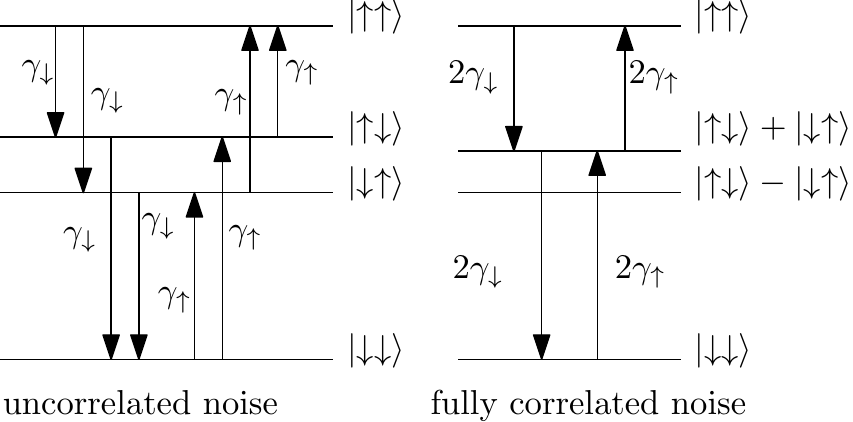}
\caption{Relaxation rates and excitation rates for a qubit pair in uncorrelated (left) and fully correlated (right) environments. For uncorrelated decoherence all states in the subspace $\{\ket{\uparrow \downarrow}, \ket{\downarrow \uparrow}\}$ decay at the same rate. For fully correlated noise there is one stationary (i.e.~decoherence-free) state and one that decays twice as fast. The rates are given by the system-bath coupling strength $\nu$ and the spectral function $\gamma_\downarrow=\nu^2 C(2\omega_q,0)$ and $\gamma_\uparrow=\nu^2 C(-2\omega_q,0)$.}
\label{fig qubit pair with correlated relaxation}
\end{figure}

In an uncorrelated environment relaxation is easily understood. In a low-temperature or vacuum environment a state with $m_{\rm exc}$ qubits in the excited state and $m_{\rm gr}$ qubits in the ground state will have $m_{\rm exc}$ transition rates\footnote{A transition rate or relaxation rate from state $\ket{a}$ to state $\ket{b}$ means that $\frac{d}{dt} \rho_{aa}= - \gamma\rho_{aa} +...$ and $\frac{d}{dt} \rho_{bb}= \gamma \rho_{aa} +...$.} into lower energy states and $m_{\rm gr}$ rates from higher energy states. In a fully correlated environment additional terms cancel or enhance certain relaxation rates\cite{McCutcheon2009}. For pairs of qubits the state $(\ket{\uparrow \downarrow}-\ket{\downarrow\uparrow})/\sqrt{2}$ is relaxation-free, i.e.~stationary, while the state $(\ket{\uparrow \downarrow}+\ket{\downarrow\uparrow})/\sqrt{2}$ decays twice as fast to the ground state as for uncorrelated decoherence. Furthermore the state $\ket{\uparrow \uparrow}$ has only one decay rate (instead of two) into the state $\ket{\uparrow \downarrow}+\ket{\downarrow\uparrow}$. This effect was mentioned in ref.~\onlinecite{Ojanen2007} and is completely analogous to Dicke's model of super- and sub-radiance in an atomic gas\cite{Dicke1954}. 

The result for two qubits does not generalize to more qubits easily. The analogy to the Dicke model, which is usually discussed in terms of the total angular momentum of the system, can be used to understand the dynamics for more qubits via the Clebsch-Gordan coefficients. For example there is the relaxation-free state $\ket{\Psi}$ with $S_z \ket{\Psi}=\sum_j \sigma_z^{(j)}\ket{\Psi}=0$ and zero total spin $S^2\ket{\Psi}= S_x^2+S_y^2+S_z^2 \ket{\Psi} = 0$ (see ref.~\onlinecite{Dicke1954}).

\subsubsection{Single excitation subspace}
In low-temperature systems the equilibrium state is very close to the ground state and the dynamics of a single excitation in a system of $n$ qubits is often of interest. For this subspace of states with only one excitation, the two qubit example gives us a good understanding of the dynamics. The subspace is spanned by the $n$ states:
\begin{align}
\begin{aligned}
	\ket{1} =&\ket{\uparrow \downarrow \downarrow \dots \downarrow}\\
	\ket{2} =&\ket{\downarrow \uparrow \downarrow \dots \downarrow}\\
	\ket{3} =&\ket{\downarrow \downarrow \uparrow \dots \downarrow}\\
&\dots\\
\ket{n} =&\ket{\downarrow \downarrow \downarrow \dots \uparrow}
\end{aligned} \label{single excitation subspace states}
\end{align}
Since all but one qubits are in the ground state we can identify $n-1$ decoherence-free, i.e.~stationary states:
\begin{align}
\begin{aligned}
	&\ket{s_j}=\nu_{j+1} \ket{1} - \nu_1 \ket{j+1} \\
\end{aligned}
\end{align}
where $\nu_j$ is the coupling strength of the $j$th spin to the environment.
Of course we could also choose any other pair, however with the given set of states we have chosen $n-1$ linearly independent (but not orthogonal) states. Further pairs would only be superpositions of the given set of stationary states. The linear independence becomes clear when we note that each stationary state is a superposition of \ket{1} with respectively one other state of the orthogonal set \eqref{single excitation subspace states}.

Since we do not regard dephasing here any superposition of decoherence-free states is also decoherence-free (or more precisely relaxation-free). In other words the stationary states span a decoherence-free subspace.

To find the one remaining state that is required to make the stationary states a basis (of the single excitation subspace) one can first orthonormalise the $n-1$ stationary states via Gram-Schmidt orthogonalisation, then start with \ket{1}, again subtract the existing orthonormal states weighted with their overlap and find the one state which is not in the decoherence free subspace. We find the general normalised form of this decaying state for a $n$-qubit system to be: 
\begin{align}
	\ket{d}=& \frac{1}{\sqrt{\sum_j \nu_j^2}} \sum_{j=1}^{n} \nu_j \ket{j}
\end{align}
One can easily see that this is the missing vector to span the one-excitation-subspace since it is orthogonal to all $n-1$ linearly independent $\ket{s_j}$ 
\begin{eqnarray}
	\forall \ j \in \left\{1, \dots ,n-1\right\} : \ \left\langle d | s_{j} \right\rangle &=& 0 \ ,
\end{eqnarray}
and the set $\left\{\ket{d},\ket{s_j} \right\}$ forms a non-orthonormalised basis of the single-excitation-subspace. As the relaxation connects the single-excitation-subspace with the zero-excitation-subspace we also define a shorthand notation for the ground-state
\begin{eqnarray}
	\ket{g} &=& | \downarrow \downarrow \downarrow \dots \rangle \\
	\mathcal{B} &=& \left\{\ket{s_j},\ket{d},\ket{g} \right\} \ .
\end{eqnarray}

In the case that the correlation length of the decoherence $\xi$ is much larger than the length of the system the contribution of the combination of the relaxation-operators on two different sites has the same weight in the Bloch-Redfield-equation as the contribution of two relaxation-operators on the same site:
\begin{eqnarray}
\lim_{\xi \rightarrow \infty}	C_\perp \left(\omega,|x_j-x_k|\right) 
	&=& C_\perp\left(\omega,0\right) 
\end{eqnarray}
We can combine the $\sigma_x$-system-coupling operators on each individual spin $\sigma_x^{(j)}$ into one coupling operator $\Sigma_x$. Only one system-coupling-operator remains and we obtain:
\begin{eqnarray}
	\Sigma_x &=& \sum_{j} \nu_j \sigma_{x}^j \\
	C_\perp^{\Sigma}(\omega) &=& C_\perp(\omega,0) \ , 
\end{eqnarray}
where $C_\perp^{\Sigma}(\omega)$ is the spectral function of the combined system-coupling-operator. Note that the spectral-function no longer depends on spatial distance since we assumed perfect correlation over the system-length. Since the $n-1$ states $\left\{s_i \right\}$ belong to the relaxation-free-subspace we only need to consider how $\Sigma_x$ operates on the rest of the basis $\mathcal{B}$. 
The $\Sigma_x$-operator also connects the single-excitation-subspace to higher excitation-subspaces. We can neglect these transitions as we assume a low-temperature environment where the transition to higher-lying energy-states due to the environmental coupling is strongly suppressed.
We find that the $\Sigma_x$-operator connects the $\ket{g}$ and $\ket{d}$-states,
\begin{eqnarray}
	\bra{g} \Sigma_x \ket{d}  &\ne& 0 \ ,
\end{eqnarray}
and opens a relaxation channel for the population in state $\ket{d}$.
Thus the single-excitation-subspace together with the ground-state separates into an $(n-1)$-dimensional decoherence-free-subspace and an effective two-level-system with a transversal noise coupling $\Sigma_x$. 

\subsubsection{Relaxation blocking by uncoupled spins}
\begin{figure}
\centering
\includegraphics{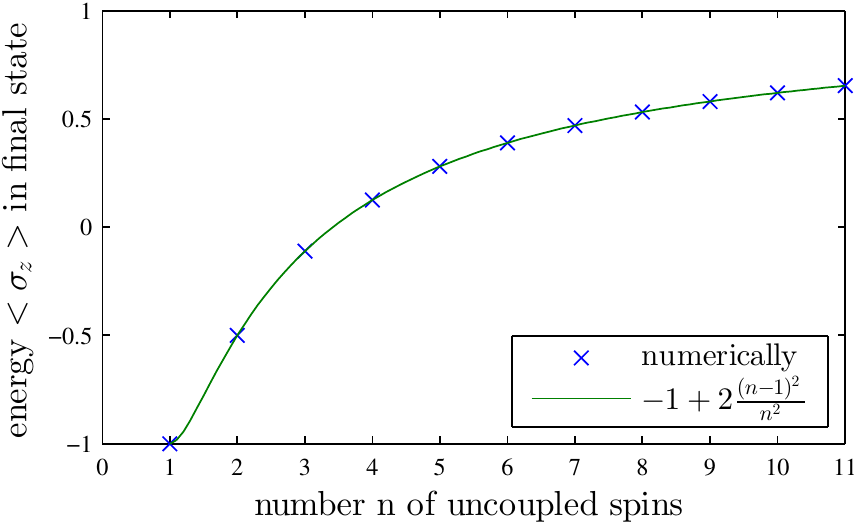}
\caption{Relaxation of a single spin, which is initially in its excited state, $\langle\sigma_z\rangle=1$. Other spins in their ground state are coupled to the same environment. Plotted is $\langle \sigma_z \rangle$ of the excited spin in the final state of the system dependent on the total number of spins. The relaxation of the spin is partially blocked because a relaxation-free subspace is formed, which overlaps more and more with the initial state as the number of spins increases.}
\label{fig relaxationblock}
\end{figure}

The existence of the stationary states $\ket{s_j}$ leads to a paradoxical effect. Uncoupled spins in their ground state reduce the relaxation of one spin in its excited state if they are all coupled to the same bath (i.e.~their noise is perfectly correlated). We investigate this phenomenon numerically by measuring the $\langle \sigma_z \rangle$ expectation value of the excited spin for very large times. We do so with an increasing number of spins in their ground state which are not coupled to the excited spin but coupled to the same environmental noise. Figure \ref{fig relaxationblock} shows that the uncoupled spins block the relaxation of the one excited spin.

The final state $\rho_f$ can be calculated analytically by dividing the initial state $\ket{1}$ into a stationary and a decaying part by projection onto the orthonormalised basis of the respective subspace.
\begin{align}
\ket{1}&=\ket{p_s}+\ket{p_d}\\
\ket{p_s} &= \ket{1} - \langle d | 1 \rangle \ket{d} \\
\ket{p_d} &=  \langle d | 1 \rangle \ket{d} \\
\rho_f&=\ket{p_s}\bra{p_s} + \braket{d}{1} \; \ket{g} \bra{g} \ ,
\end{align}
with
\begin{eqnarray}
	\braket{d}{1} &=& \frac{1}{\sqrt{\sum_j \nu_j^2}} \nu_1 
\end{eqnarray}
we then find the energy $\langle \sigma_z^{(1)} (t \rightarrow \infty)\rangle=\Tr( \sigma_z^{(1)} \rho_f )$ of the initially excited qubit in the final state:
\begin{eqnarray}
\langle \sigma_z^{(1)} (t \rightarrow \infty) \rangle &=& \langle 1| \sigma_z^{(1)}  | 1 \rangle - 2 \braket{d}{1}  \langle d| \sigma_z^{(1)} |1 \rangle  \nonumber \\
&+& \braket{d}{1}^2  \langle d | \sigma_z^{(1)} |d \rangle + \braket{d}{1} ^2  \langle g | \sigma_z^{(1)} | g \rangle \nonumber \\
&=& -1 + \frac{2 \left( \sum_{j=2}^{n} \nu_j^2 \right)^2}{\left(\sum_{j=1}^n \nu_j^2\right)^2} 
\end{eqnarray}
where again $\nu_j$ is the coupling of the $j$th spin to the environment. When the coupling of all spins to the environment is equally strong (as assumed in our numerics) the expression simplifies to:
\begin{align}
\langle \sigma_z^{(1)} (t \rightarrow \infty) \rangle = -1 + 2\frac{(n-1)^2}{n^2} \ ,
\end{align}
This generalisation for $n$ spins compares to the numerical calculations very well as can be seen in figure \ref{fig relaxationblock}.
The total energy remaining in the system after total relaxation can be measured by the total $S_z = \sum_{j} \sigma_z^{(j)}$ operator. For better comparability we subtract the ground state energy of $-n$. The general expression is
\begin{eqnarray}
	\langle S_z\left( t\rightarrow \infty\right) \rangle + n &=& 2 -\frac{2\nu_1^2}{\sum_{j=1}^n \nu_j^2} 
\end{eqnarray}
For all $\nu_j$ equal we find:
\begin{align}
\langle S_z\left( t\rightarrow \infty\right) \rangle + n &= 2- \frac{2}{n} 
\end{align}
The total excitation remaining in the system after total relaxation is distributed over all sites. Even though there is no coherent coupling between neighbouring qubits, the spatially correlated relaxation leads to an excitation-transfer between the qubits. The fraction of the excitation transferred $\sigma_{\rm trf} $ from the first into the other qubits is reduced with increasing system-size $n$:
\begin{eqnarray}
	\sigma_{\rm trf} &=& \frac{2}{n}-\frac{2}{n^2}  \ .
\end{eqnarray}

To summarise, when a number of uncoupled spins are subject to the same environmental noise, then the single excitation subspace is predominantly relaxation-free and contains only one decaying state. As a consequence we find a general effect: a spin's relaxation can always be blocked by other spins in their ground state, which are not coupled to the excited spin but only to the same environmental noise.

\section{Transfer in spin chains}
\label{sec transfer in spin chains}
We will now regard a particular set-up for perfect state transfer through a spin chain and present the effects of spatially correlated decoherence on the transfer. 

\begin{figure}
\centering
\includegraphics[scale=0.96]{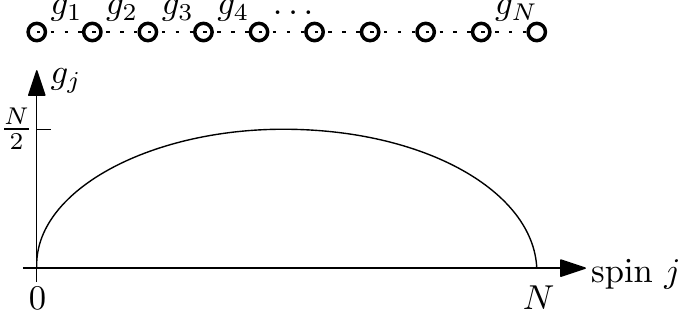}
\caption{The coupling strength $g_j=g\sqrt{j(N-j)}$ between spins describes a half circle, i.e.~is strongest in the middle of the chain, to guarantee perfect state transfer. For all numerical simulations we used the following parameters: $\omega_q=100,\;g_j=\sqrt{j(N-j)},\;v_j=0$ or $1,\;\nu_j=0$ or $1$.} 
\label{fig numerical parameters}
\end{figure}

\subsection{The model system}
\label{sec the model system}
\begin{figure}
\centering
\includegraphics[scale=1]{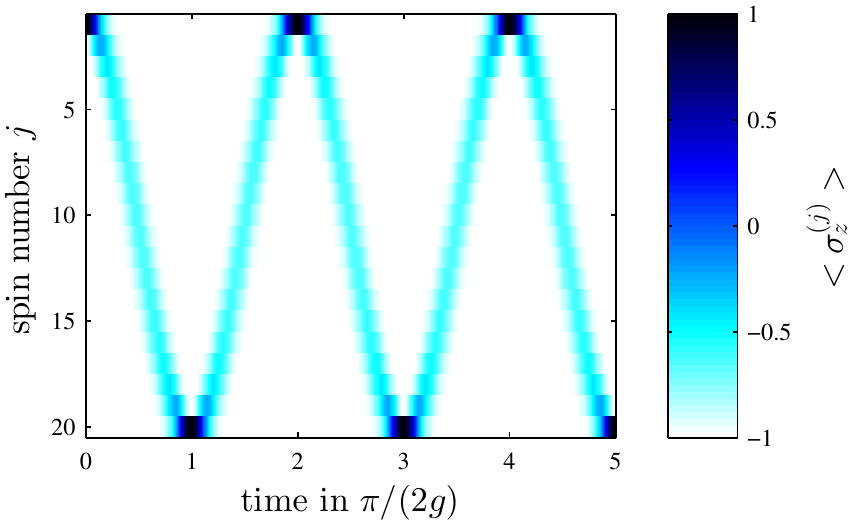}
\caption{Coherent dynamics of the given spin chain, eq.~\eqref{spin chain Hamiltonian}. The excitation is transferred from one end of the spin chain to the other. This mechanism depends on the coherence of the spins because the excitation spreads out before it refocuses at the other end. This is an ideal model system to test the influence of decoherence with different spatial correlation lengths on the excitation transfer.}
\label{fig spin chain state transfer}
\end{figure}

The system is a linear chain  of $N$ transversely coupled spins:
\begin{align}
H_s = \sum_{j=1}^N \omega_q \sigma_z^{(j)} + \sum_{j=1}^{N-1} \frac{g_j}{2} \left(\sigma_x^{(j)} \sigma_x^{(j+1)} + \sigma_y^{(j)} \sigma_y^{(j+1)}\right) \label{spin chain Hamiltonian}
\end{align}
with the level splitting $\omega_q$ and the coupling strength $g_j$ of spin $j$ to its right neighbour. To guarantee perfect state transfer\cite{Christandl2004} the coupling strength is chosen to have the form $g_j=g \sqrt{j(N-j)}$. Note that the coupling strength plotted as a function of the position $j$ describes a half circle (figure \ref{fig numerical parameters}). Initially we choose the spin at the start of the chain to be excited and all other spins in the ground state, i.e.~$\ket{\uparrow \downarrow \downarrow \downarrow \dots}$. The coherent dynamics of this system is shown in figure \ref{fig spin chain state transfer}. The excitation, initially at one end of the chain spreads out, travels through the chain and, due to the particular profile of the coupling strength, refocuses at the other end. Due to the symmetry of the system that process then starts again in reverse. The time it takes for the excitation to pass through the chain once is $\pi/(2g)$. This system is an ideal model system to test the influence of environmental noise with different correlation lengths on excitation transfer because it has a clearly defined end point of the transfer, while the transfer process depends on the coherence of the spins. As in section \ref{sec spatially correlated effects}, we discuss longitudinal and transversal bath couplings separately.

\subsection{Dephasing}
First we regard longitudinal coupling to the environment:
\begin{align}
H_{int}= \sum_{j=1}^N v_j\,\sigma_z^{(j)}\, B_\parallel^{(j)} 
\end{align}
where $B_\parallel^{(j)}$ is a bath operator and $v_j$ the bath coupling strength. We assume a Gau\ss ian shape of the spatial correlation function\cite{Jeske2013formalism} associated with $B_\parallel^{(j)}$:
\begin{align}
C(\omega=0,|x_j-x_k|)= 
2^{-\frac{(x_j-x_k)^2}{\xi^2}}
\end{align}
with the correlation length $\xi$.

\begin{figure}
\centering
\includegraphics[scale=1]{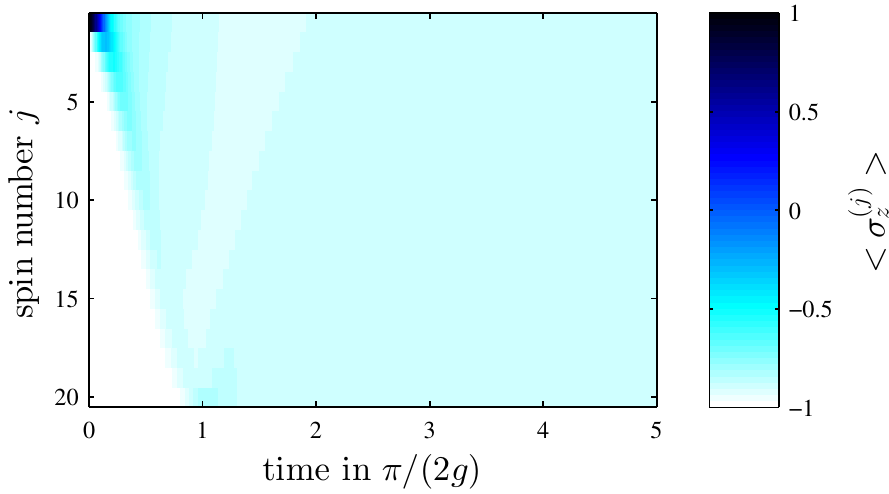}
\includegraphics[scale=1]{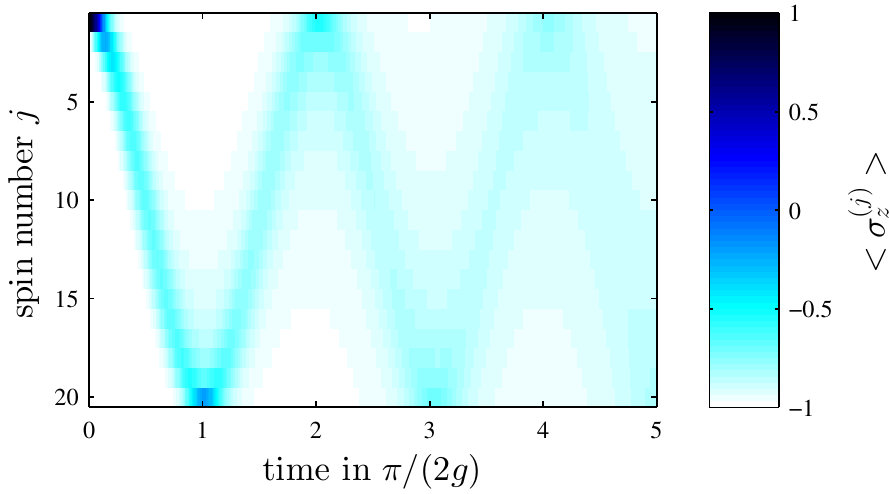}
\includegraphics[scale=1]{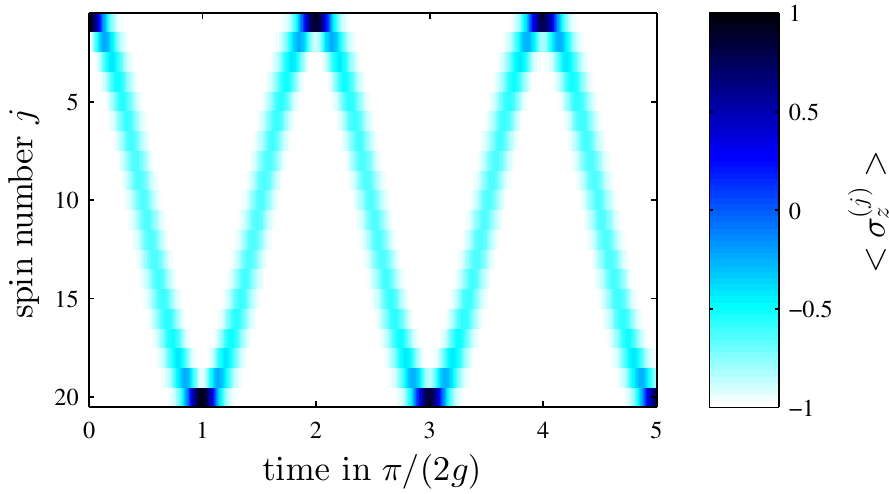}
\caption{Dynamics of the spin chain with constant longitudinal system-bath coupling $v_j=1$, and different correlation lengths, \textbf{top}: $\xi=0.2d$, \textbf{middle}: $\xi=2d$, \textbf{bottom}: $\xi=20d$ where $d$ is the distance between spins. The relatively strong environmental coupling leads to dephasing, which for uncorrelated decoherence (top) makes the excitation spread out over the chain and destroys the transfer. With increasing correlation length of the environment the coherent dynamics (cf. figure \ref{fig spin chain state transfer}) is restored even though the system-bath coupling is not decreased.}
\label{fig dephasing in spin chain}
\end{figure}

As the dynamics of the excitation transfer depends on the coherence of the spins, uncorrelated dephasing destroys the refocusing at each end and spreads the excitation out over the whole chain (figure \ref{fig dephasing in spin chain}, top). With increasing correlation length $\xi$ the detrimental influence of the environment is reduced and the excitation transfer is restored without a change in the noise strength $v_j$ (figure \ref{fig dephasing in spin chain}). The dynamics of the transfer relies on the coherence of the single excitation subspace, which approaches a decoherence-free subspace for long correlations $\xi \rightarrow \infty$ (section \ref{sec spatially correlated dephasing}). 

\begin{figure}
\centering
\includegraphics[scale=1]{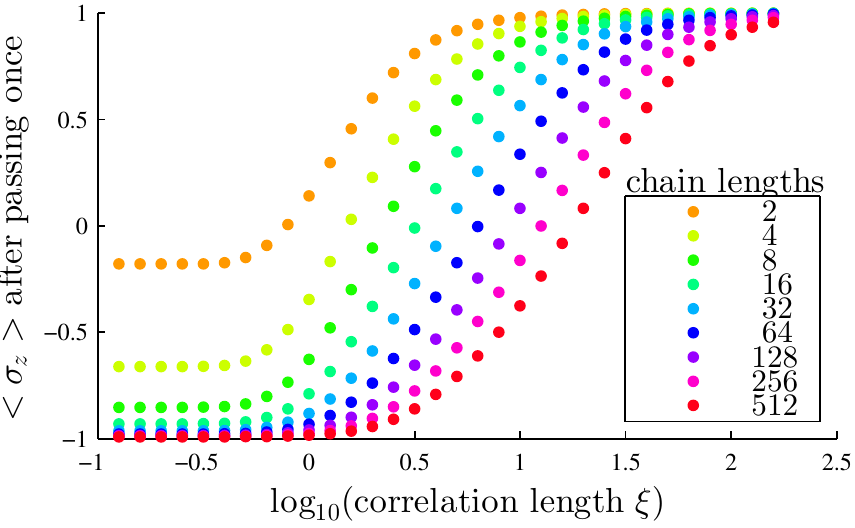}
\caption{Transfer quality of an excitation through the spin chain dependent on the correlation length for different chain lengths (see legend). The characteristic ``step'' in the transfer quality defines a critical correlation length $\xi_c$, which only changes with chain length and is independent of other parameters. }
\label{fig dephasingstep in spin chain}
\end{figure}

To quantify the excitation transfer, we measure $\langle\sigma_z\rangle$ of the end spin after one passing through the chain at $t=\pi/(2g)$. We plot this result dependent on the correlation length $\xi$ in figure \ref{fig dephasingstep in spin chain} and find a clear step in the transfer quality, which means there is a particular critical correlation length $\xi_c$. Noise with a correlation length below $\xi_c$ destroys the transfer, while above $\xi_c$ the quality of the transfer is high. Numerical results show that the critical correlation length does not depend on the noise intensity $v_j$ in the weak coupling regime $v_j < \min(g_j)$. 

The major determining influence is the chain length (figure \ref{fig dephasingstep in spin chain}). For the perfect state-transfer protocol, increasing chain length also increases the maximum spread (or packet width) of one excitation in the transfer, which occurs at half the passing time $t=\pi/(4g)$ (cf.~fig.~\ref{fig spin chain state transfer}). In other words the critical noise correlation length $\xi_c$ depends on the maximal packet width in the chain, which is an intuitive result as the transfer depends on the refocusing of that excitation packet. To quantify this statement we determined both quantities numerically for the different chain lengths given in figure \ref{fig dephasingstep in spin chain} and confirmed a linear relationship: the positions $\xi_c$ of maximal gradient in figure \ref{fig dephasingstep in spin chain} depends as $\xi_c \approx 1.7w_p-0.89$ on the half width at half maximum of the excitation packet $w_p$ after $t=\pi/(4g)$. 

The linear dependence of $\xi_c$ on the chain length suggests, that excitation transfer is not impaired by noise as long as the noise is correlated on a length scale that goes beyond the maximal packet width of the excitation. Similarly, the dynamics of a single excitation in a spin network in general is not impaired by noise that is correlated on a larger scale than the spread of the excitation. 

\begin{figure}
\centering
\includegraphics[scale=1]{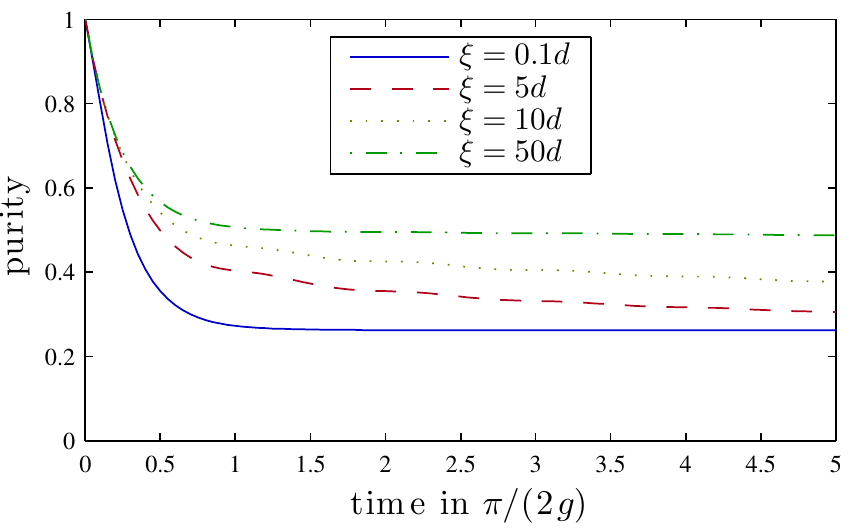}
\caption{Purity $\Tr(\rho^2)$ as a function of time with dephasing noise for different correlation lengths. The purity is partially restored for longer correlation lengths, however there is a residual decay because the phase coherence to the ground state is lost, despite the fact that excitation transfer is fully restored for long correlation lengths (cf.~fig.~\ref{fig dephasing in spin chain}).}
\label{fig dephasing in spin chain sx}
\end{figure}

The phase coherence to the ground state decays regardless of the correlation length $\xi$ even when the excitation transfer is restored because the ground state has a difference of one excitation to the single-excitation subspace. This can be seen when the purity $\Tr(\rho^2)$ is measured (figure \ref{fig dephasing in spin chain sx}).  This loss of the phase information in the given set-up means that for spatially correlated noise the excitation transfer is no longer a \emph{state} transfer in the sense of quantum information but has become a classical bit transfer. One way that this problem might be overcome is via a Hahn echo technique, where a $\pi/2$ bit flip to the entire chain is incorporated after half of the transfer time. However, this would be a more technologically challenging set-up. Outside quantum information there are applications in which the excitation transfer with ``classical information'' is equally desirable, e.g.~photosynthetic systems. In these situations correlated dephasing enables the transfer at high qualities even for relatively strong noise.

\begin{figure}
\centering
\includegraphics[scale=1]{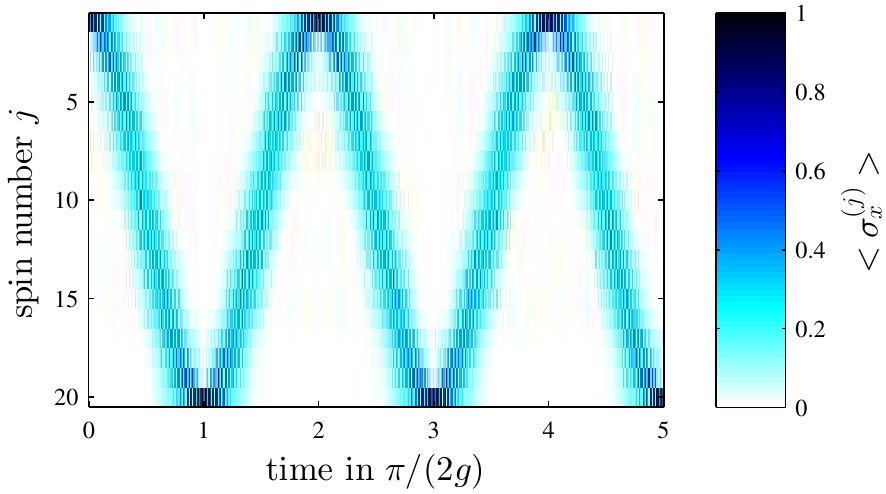}
\includegraphics[scale=1]{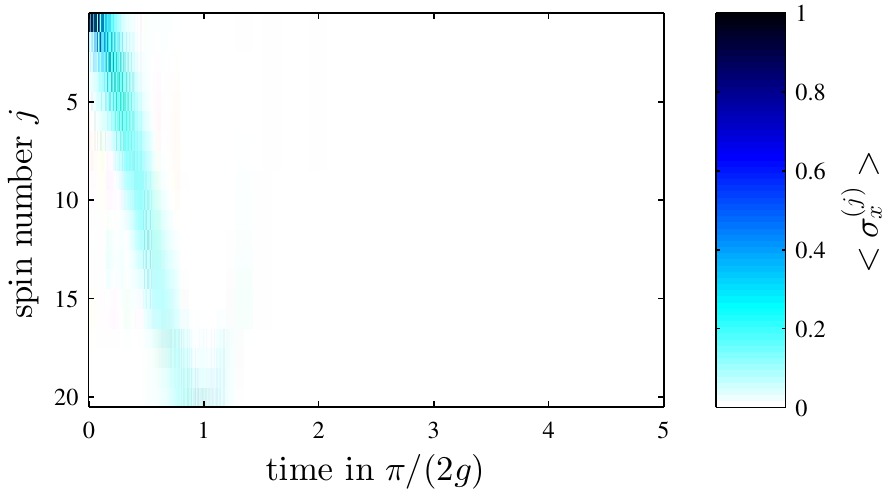}
\caption{\textbf{Top}: Coherent evolution of the system where the magnitude of the transversal magnetisation $|\langle \sigma_x \rangle|$ is plotted. The $\langle \sigma_x \rangle$ expectation value has a high frequency harmonic ($\propto 2\omega_q$) but here we plot $|\langle \sigma_x \rangle|$ to focus on the slow dynamics. The other spins in their ground state have zero expectation value. \textbf{Bottom}: In decoherent evolution, the phase information is lost very quickly even for broadly correlated noise (here: $\xi=20d$). Note that the bottom plot is the \emph{same} evolution as the bottom plot of figure \ref{fig dephasing in spin chain}, i.e.~the excitation is transferred very well, but the phase coherence to the ground state is lost very quickly at the same time.}
\end{figure}

\subsection{Relaxation}

\begin{figure}
\centering
\includegraphics[scale=1]{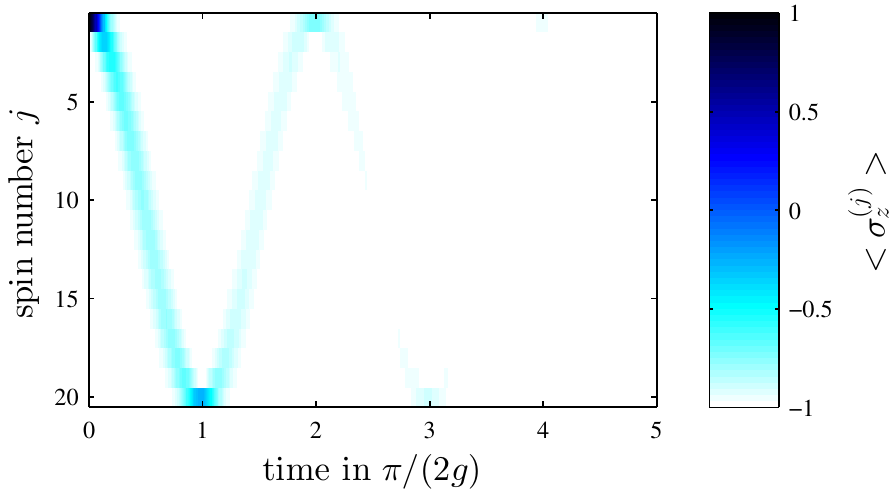}
\includegraphics[scale=1]{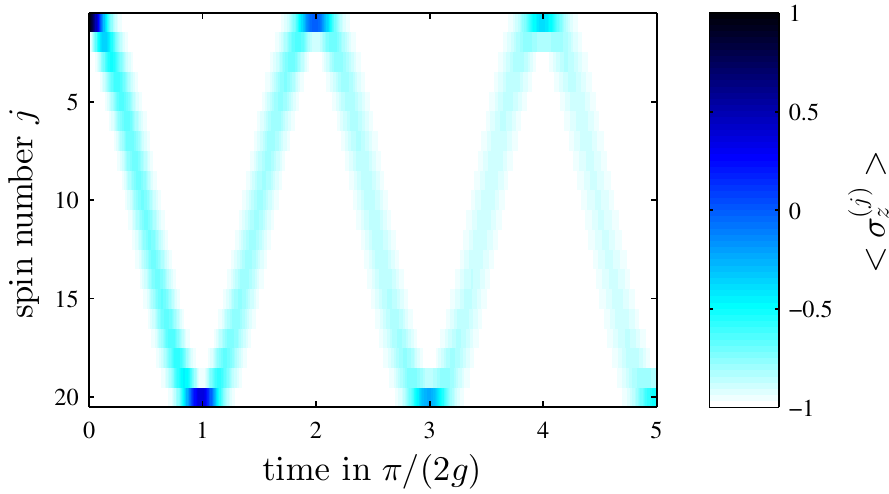}
\includegraphics[scale=1]{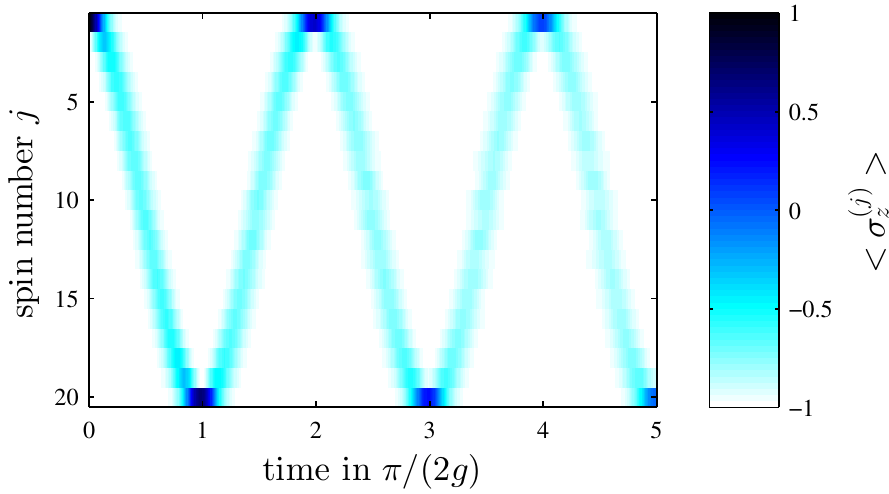}
\caption{Relaxation (due to transversal coupling) with different correlation lengths: \textbf{top}: $\xi=0.2d$, \textbf{middle}: $\xi=2d$, \textbf{bottom}: $\xi=20d$ where $d$ is the distance between spins. Similar to dephasing (figure \ref{fig dephasing in spin chain}) long correlation lengths are advantageous for the transfer quality.}
\label{fig spin chain relaxation}
\end{figure}

\begin{figure}
\centering
\includegraphics[scale=1]{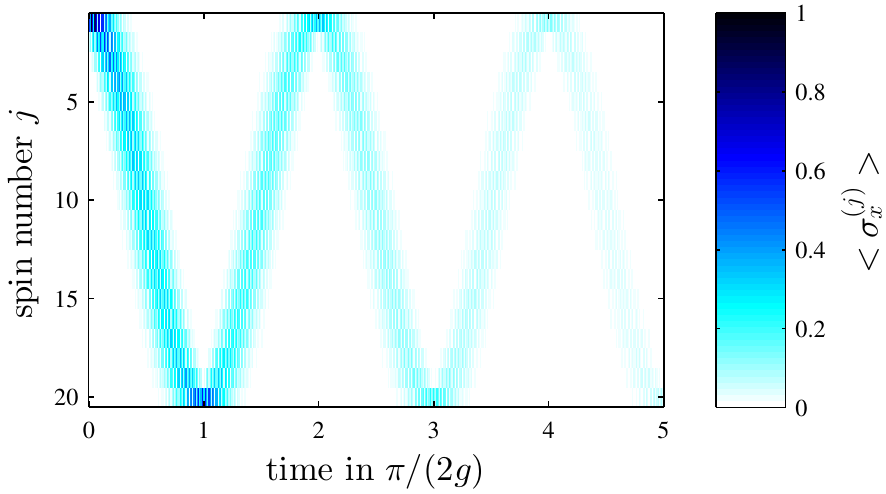}
\includegraphics[scale=1]{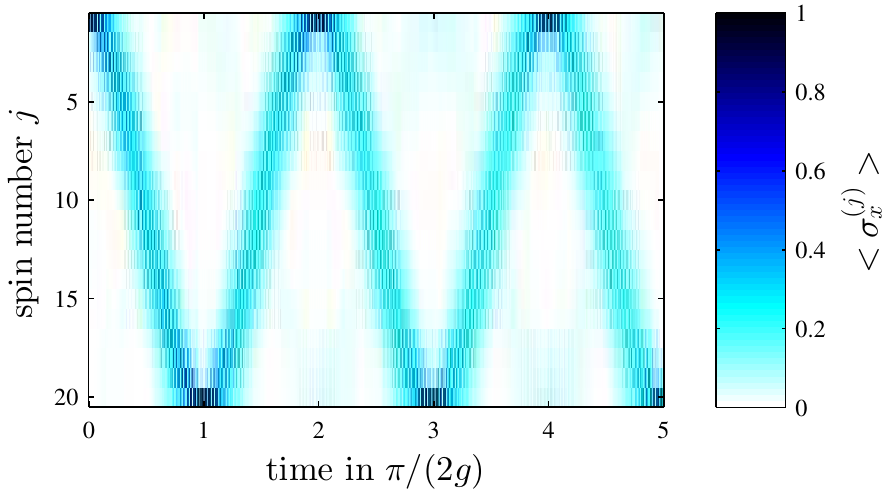}
\caption{Evolution of the system undergoing pure relaxation, where the transversal magnetisation $|\langle \sigma_x\rangle|$ is plotted. \textbf{top}: $\xi=0.2d$; \textbf{bottom}:$\xi=20d$. For long correlation length the phase information is also preserved and transferred through the spin chain.}
\label{fig relaxation in spin chains sx}
\end{figure}

In this section we will discuss the effects of transversal bath coupling. Note that a combined appearance of both longitudinal and transversal couplings does not alter any of the effects described in this section but merely adds dephasing as discussed above.

The spin chain with Hamiltonian \eqref{spin chain Hamiltonian} is now coupled transversely to the environment:
\begin{align}
H_{int}= \sum_{j=1}^N  \nu_j\,\sigma_x^{(j)}\, B_\perp^{(j)} 
\end{align}
with the coupling strength $\nu_j$. We assume a vacuum or low temperature environment, i.e.~the spectral function at negative frequency $-\omega_q$ is approximately zero. For positive frequency $\omega_q$ we again assume Gau\ss ian shaped spatial correlations which are constant in frequency:
\begin{align}
C(-\omega_q, |x_j-x_k|) &= 0\\
C(\omega_q,|x_j-x_k|) &= 2^{-\frac{(x_j-x_k)^2}{\xi^2}}
\end{align}
This means we will only find energy loss from the spin chain and no excitation gain from the environment will occur. 

\begin{figure}
\centering
\includegraphics[scale=1]{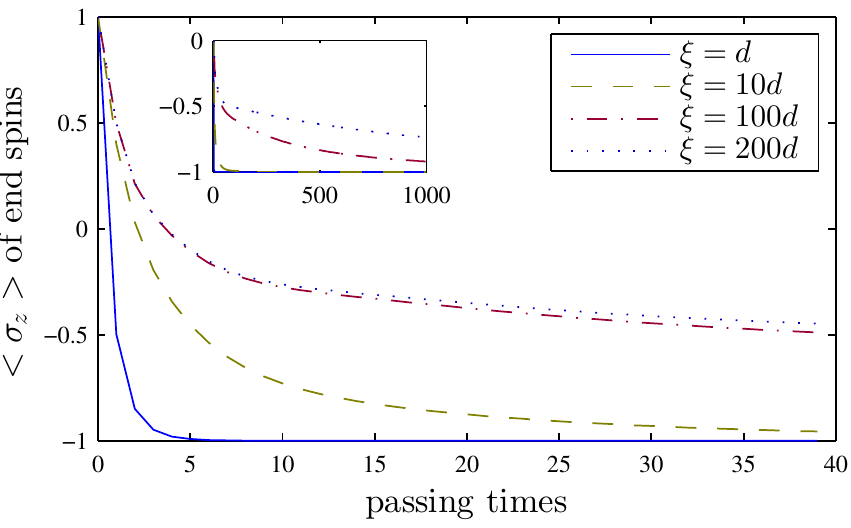}
\caption{Transfer quality for continuing evolution with several different correlation lengths. For long correlation lengths two separate decay time scales arise.}
\label{fig spin chains two time scales relaxation}
\end{figure}

\begin{figure}
\centering
\includegraphics[scale=1]{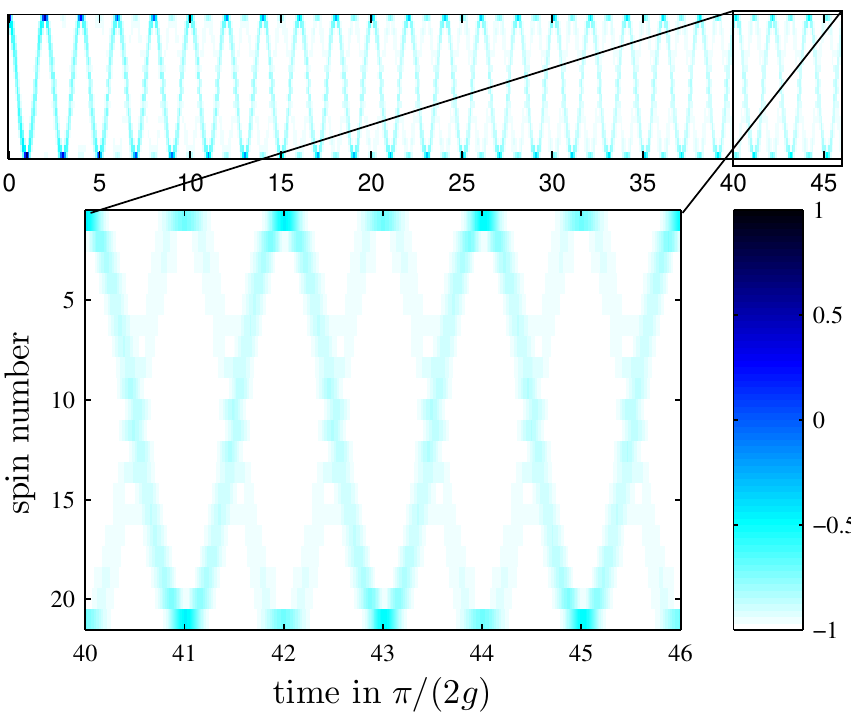}
\caption{$\langle \sigma_z^{(j)} \rangle$ expectation value after 40 passings through the chain with relaxation and long correlation length $\xi=100d$. The dynamics shows the intermediate state, eq.~\eqref{spin chain intermediate state}, which occurs after the fast decay and before the slow decay in figure \ref{fig spin chains two time scales relaxation}. }
\label{fig spinchain intermediate state}
\end{figure}

In the time evolution we find that longer correlation lengths $\xi$ are advantageous for the transfer quality (Figure \ref{fig spin chain relaxation}). This is similar to dephasing. However, contrary to dephasing the phase information is also preserved for longer correlation lengths $\xi$ (figure \ref{fig relaxation in spin chains sx}). Furthermore the relaxation is not simply slowed down for long correlations. Two relaxation time scales emerge, a fast one and a slow one. This can be visualised by taking only the time points which are multiples of $\pi/2$, where in the coherent dynamics the state should be refocused at the ends of the chain. Plotting the expectation value of the respective spin displays a continuous decay with two distinct regimes. Figure \ref{fig spin chains two time scales relaxation} shows clearly the two separate time scales. If we consider the dynamics of the spin chain at the time after the fast decay has finished, and the slow decay is just starting, we find that the excitation, which started initially at one end, is now split up and refocuses at both ends simultaneously (figure \ref{fig spinchain intermediate state}). The corresponding density matrix at these points in time is given by a statistical mixture of two states:

\begin{align}
\ket{\Psi_1} &=(\ket{\uparrow \downarrow \downarrow\dots \downarrow}+\ket{\downarrow\downarrow\downarrow\dots\downarrow\uparrow})/\sqrt{2} & p_1 &= 0.5 \label{spin chain intermediate state}\\
\ket{\Psi_2} &= \ket{\downarrow \downarrow \downarrow \dots \downarrow} & p_2 &= 0.5 \label{spin chain relaxation intermediate state}
\end{align}
In other words the relaxation has entangled the first spin and the last spin with an efficiency of 50\%. This entangled state then decays on a much slower time scale.

Again we can explain the behaviour with the analytical results for perfectly correlated environments from section \ref{sec spatially correlated relaxation}. The coherent dynamics is entirely in the single-excitation subspace, which consists of $n$ states. For perfect correlations $\xi \rightarrow \infty$ the single-excitation subspace contains only one decaying state and a relaxation-free (or subradiant) subspace of $n-1$ states. The coherent dynamics of the chain moves the excitation around and transfers probability between the relaxation-free states and the decaying state. All population in the decaying state however relaxes into the ground state on the short time scale. The only exception is the state $\ket{\Psi_1}$, which has a measurement probability in the initial state of $|\braket{\Psi_1}{\uparrow \downarrow \downarrow \dots}|^2=1/2$. Figure \ref{fig spinchain intermediate state} shows that in this state the excitations on both ends travel through the chain simultaneously. In other words the state $\ket{\Psi_1}$ evolves almost entirely in the relaxation-free subspace into itself, resulting in a much slower decay rate.

\section{Conclusions}
We studied general effects for environmental noise with long spatial correlation lengths in systems of several spins. While for short correlation lengths the dephasing rate between two states is proportional to the number $n_f$ of flipped spins one finds that for long correlation lengths it becomes proportional to $n_e^2$, where $n_e$ is the difference in the number of excitations. This leads to much stronger dephasing between certain states but also the creation of dephasing-free subspaces. For relaxation the dynamics becomes rather complex for long correlation lengths. Characteristic is the fact that the mixed terms in the master equation, which involve operators acting on two different spins, cancel or enhance certain relaxation rates, dependent on the state of the respective pair. For a pair of spins in the state $(\ket{\uparrow \downarrow} + \ket{\downarrow \uparrow})/\sqrt{2}$ the rate is twice as high as for uncorrelated noise. In contrast relaxation rates are cancelled for pairs in the state $(\ket{\uparrow \downarrow} - \ket{\downarrow \uparrow})/\sqrt{2}$. In the single excitation subspaces all of these $n-1$ states are therefore relaxation-free. This leads to the paradoxical effect that a qubit's relaxation can be blocked by coupling other qubits, which are in their respective ground state, to the same noise environment.

For excitation transfer, spatially correlated noise is strongly advantageous compared to uncorrelated noise. The detrimental effects of dephasing on the transfer dynamics vanish as the noise correlation length is greater than the maximal packet width of the excitation in the transfer. The excitation can then be transferred with very high fidelity even for strong noise. While the dynamics of the transfer is restored with long correlation length the phase coherence to the ground state is still lost in the transfer and the high-fidelity excitation transfer is no longer a perfect \emph{state} transfer. 

With relaxation the transfer also improves with increasing spatial correlation length of the noise. The relaxation time increases and two separate time scales arise. Initially the state $\ket{\uparrow \downarrow \downarrow \dots\downarrow}$ relaxes into the entangled state $(\ket{\uparrow \downarrow \downarrow \dots \downarrow} - \ket{\downarrow \downarrow \dots \downarrow \uparrow})/\sqrt{2}$. This intermediate state is very robust and decays on a longer time scale. It can be concluded that spatially correlated noise displays significantly different dynamics to spatially uncorrelated noise. Longer correlation lengths are generally advantageous to quantum transport as they reduce dephasing effects  and produce an intermediate entangled state with reduced relaxation rates.

\acknowledgments
N.V. acknowledges support from the Deutscher Akademischer Austausch Dienst (DAAD). 

\bibliographystyle{apsrev4-1}
\bibliography{publication}
\end{document}